\documentclass[aps,pra,superscriptaddress,tightenlines,showpacs,floatfix,notitlepage,amsmath,amssymb]{revtex4-1}

\usepackage{physics}
\usepackage{mathtools}
\usepackage{graphicx}
\usepackage{bm}
\usepackage{amssymb,amsmath,latexsym,mathrsfs}
\usepackage[usenames,dvipsnames]{color}
\usepackage[breaklinks,colorlinks,urlcolor=purple,citecolor=purple,linkcolor=purple]{hyperref}
\usepackage{float}
\usepackage{xcolor}
\usepackage{units}

\begin{document}

\title{A Resolution of the Ito-Stratonovich Debate in Quantum Stochastic Processes}

\author{Aritro Mukherjee}

\affiliation{Faculty of Physics, University of Duisburg-Essen, Lotharstraße 1, 47048 Duisburg, Germany}

\begin{abstract}
Quantum stochastic processes are widely used in describing open quantum systems and in the context of quantum foundations. Physically relevant quantum stochastic processes driven by multiplicative colored noise are generically non-Markovian and analytically intractable. Further, their Markovian limits are generically inequivalent when using either the Ito or Stratonovich conventions for the same quantum stochastic processes. We introduce a quantum noise homogenization scheme that temporally coarse-grains non-Markovian, colored-noise-driven quantum stochastic processes and connects them to their effective white-noise (Markovian) limits. Our approach uses a novel phase-space augmentation that maps the non-Markovian dynamics into a higher-dimensional Markovian system and then applies a controlled perturbative coarse-graining scheme in the characteristic time scales of the noise. This allows an explicit analytical algorithm to derive effective Markovian generators with renormalized coefficients and enables various physical constraints, such as causality, to be imposed on them. We thus resolve the Ito--Stratonovich ambiguity for multiplicative colored-noise-driven quantum stochastic processes, wherein we show that their consistent Markovian limit corresponds to the Stratonovich convention with renormalized coefficients as well \textcolor{black}{as correction terms in Ito's convention}. By assuming their Markovian limit unravels causal, completely positive and trace-preserving dynamics, we further characterize a physically relevant family of non-Markovian quantum stochastic processes driven by multiplicative colored~noise.\end{abstract}

\maketitle

\section{Introduction}
Quantum stochastic processes are random processes describing the evolution of normalized rays or states, $\ket{\psi}$, on~a complex Hilbert space, $\mathcal{H}$. They are utilized to model a wide variety of quantum systems with decoherence or dissipation or indeed, non-unitary modifications~\cite{hudson1984quantum,Parthasarathy1992,percival1998,Gisin84,QFDR2,GardinerZoller2004,Barchielli2009,aritro_PhD}. In~the non-relativistic regime, the~quantum dynamics of an isolated system is given by the unitary, deterministic and time-reversal symmetric Schr\"odinger equation $i\hbar\frac{\partial}{\partial t}\ket{\psi}=\hat{H}\ket{\psi}$ where $\hat{H}$ is the Hamiltonian~\cite{Sakurai94}. Thus, the dynamics do not admit any further randomness, unless~the system is measured explicitly or monitored indirectly. 
In~this respect, quantum stochastic processes present an advantageous, trajectory-based description of quantum randomness during monitoring and decoherence in the context of open quantum systems~\cite{Barchielli2009,GardinerZoller2004,Araujo_2019_QFDR1,Breur_Petr02,OQS_Dyn_SemiGroups_AlickiLendi2007}, as~well as in the context of foundational objective collapse theories~\cite{Bassi_03_PhyRep,Ghirardi_1986,Diosi_87_PLA,Pearle_89_PRA,Ghirardi_90_PRA,aritro1,aritro2,aritro3,aritro_SUI,gao_2018,percival1998,aritro_PhD}. These advantages may be fully exploited to analytically show that in the Markovian regime, completely positive trace-preserving (CPTP) maps such as the well known Gorini--Kossakowsky--Sudarshan--Lindblad (GKSL) dynamical semi-group~\cite{GKS76,Lindblad1976,OQS_Dyn_SemiGroups_AlickiLendi2007} generators may be unraveled in terms of quantum stochastic processes driven by white, i.e.,~temporally uncorrelated,  noise~\cite{Gisin84,percival1998,GardinerZoller2004,Barchielli2009}. In~this article, we focus on this class of completely positive and norm-preserving quantum stochastic processes, whose ensemble-averaged dynamics result in CPTP maps. It is worth noting, however, that CPTP dynamics do not encompass all physically relevant evolutions in open quantum systems (see discussions in Refs.~\cite{NoCP_Pechukas1994,NoCP_Alicki1995,NoCP_Pechukas1995Reply,NoCP_ShajiSudarshan2005,NoCP_CoppolaDaoumaHenkel2025}), a~fact we do not pursue further in this~article.
In the non-Markovian regime, specifically, in scenarios where the driving noise is temporally correlated or colored, analytically exact results are rare for non-trivial scenarios and necessitate various approximations (see Refs.~\cite{Hanggi94,Luczka2005}) to treat the same problem using the available robust Markovian techniques, which further lead to the so-called Ito--Stratonovich debate~\cite{Ito_strat1,Pavliotis2008,gardiner2004handbook}. Concretely, a~naive strategy, wherein the colored noise in a non-Markovian dynamical system is substituted by a Markovian white noise, results in inequivalent realizations of stochastic processes while employing either the Ito or Stratonovich conventions, or indeed any other convention~\cite{Pavliotis2008,Bo_whirlProd_2013,HorsthemkeBook2006}. Thus, there is an urgent need to clarify this ambiguity as well as to provide a robust prescription that can concretely map a particular physical scenario to a particular stochastic convention (Ito or Stratonovich) so as to realize effective Markovian processes corresponding to their non-Markovian~counterparts. 

In this article, we analyze this problem in a general, yet physically relevant, Hermitian or real setting and provide such a prescription by constructing a perturbative methodology to coarse-grain non-Markovian quantum stochastic processes and connect them to their Markovian limits, while explicitly clarifying when a certain stochastic convention is applicable. Our analysis shows that the Stratonovich convention naturally emerges for a class of non-Markovian quantum stochastic processes driven by colored (temporally correlated) noise, and~thus yields corrections in Ito's convention. By~further assuming that their Markovian limits unravel causal, CPTP dynamics, we uncover a family of physically relevant quantum stochastic processes that are non-Markovian and, in general, may be driven by multiplicative colored noise. This resolves the ambiguity as to which convention is physically admissible and must be chosen. Our results are expected to be applicable quite generally to non-Markovian quantum stochastic processes, encountered in open quantum systems and continuous monitoring~\cite{Breur_Petr02,Barchielli2009,GardinerZoller2004,OQS_Dyn_SemiGroups_AlickiLendi2007}, non-equilibrium many-body quantum theory~\cite{GKS76,Lindblad1976,Breur_Petr02,Araujo_2019_QFDR1,QFDR2,aritro_PhD}, the~emerging field of noise-driven phase transitions~\cite{HorsthemkeBook2006,aritro2,aritro3,aritro_PhD}, and in the modifications of quantum theory, employed in the foundations of physics~\cite{Bassi_03_PhyRep,Ghirardi_1986,Diosi_87_PLA,Pearle_89_PRA,Ghirardi_90_PRA,aritro1,aritro2,aritro3,aritro_SUI,gao_2018,percival1998,aritro_PhD}.  

\section{Quantum Stochastic~Processes}\label{sec:2}
Quantum stochastic processes constitute non-unitary extensions of Schr\"odinger's equations with additional stochastic driving, often called quantum noise~\cite{GardinerZoller2004,Barchielli2009,percival1998,Gisin84,aritro_PhD}. Typically, in~all applications, the~major object of study are non-unitary, time-dependent, and stochastic operators of the form $i\hat{G}(\xi,\psi)$. The~stochastic character of such operators are reflected by their dependence on stochastic variables $\xi$ (further explained below) as well as, crucially, on~$\ket{\psi}$, the~quantum state of the entire isolated system under study. Such operators are employed as non-unitary modifications to operator expectation values in the Heisenberg picture~\cite{Araujo_2019_QFDR1,QFDR2,hudson1984quantum,Parthasarathy1992} or employed similarly in the Schr\"odinger picture to evolve states~\cite{Gisin84,Bassi_03_PhyRep,aritro2,aritro3,percival1998,aritro_PhD}, for~example, via modified Schr\"odinger equations, $i\hbar\frac{\partial}{\partial t}\ket{\psi}=\hat{H}\ket{\psi}+i\hat{G}(\xi,\psi)$. 
For instance, in the context of open quantum systems and continuous monitoring, in~general, the~stochasticity in the dynamics of open systems is controlled by the environment, which is then further approximated based on the information one wishes to extract~\cite{Barchielli2009,GardinerZoller2004,Breur_Petr02}. This is also true in the context of constructing quantum Langevin equations for open system observables, where further constraints on the noise follow from constraints imposed on the environment~\cite{Araujo_2019_QFDR1,QFDR2}. Finally, in~the context of the foundations of quantum physics and modifications of quantum theory, the~stochasticity needs to be non-linearly dependent on the state, so as to retain normalization and avoid superluminal signaling~\cite{Gisin:1989sx,Bassi2015,aritro_FTL, aritro_PhD}. Generically, these constraints may render the equations of motion non-Markovian with multiplicative noise in both scenarios where one treats certain operators in the Heisenberg picture with noise (such as in quantum Langevin equations~\cite{Araujo_2019_QFDR1,QFDR2}), or~when one treats the time-dependent state, $\ket{\psi}\in \mathcal{H}$, in~the Schr\"odinger picture via stochastic Schr\"odinger equations~\cite{Gisin84,percival1998,Barchielli2009,GardinerZoller2004}.

\textbf{Preliminary Considerations:} 
 The above discussions imply that any physically motivated analysis requires a careful treatment of the generically non-linear ($\ket{\psi}$-dependent) and stochastic  non-unitary operators $i\hat{G}(\xi,\psi)$. Incorporating such operators to modify the time evolution generator, the~self adjoint Hamiltonian, introduces various domain complications, which we avoid by restricting our analysis to finite dimensions. Note that in~the case of infinite-dimensional separable Hilbert spaces, added assumptions are required such that all domains are compatible~\cite{BC_Hall_2013Quantum_for_math}.  We thus focus on scenarios where the dimension of the Hilbert space (which is endowed with stochastic dynamics) is finite. Note that this does not imply any further restriction on sub-Hilbert spaces such as an environment in the context of open quantum systems. In~finite dimensions, all Hermitian operators are self-adjoint and we further restrict our analysis to scenarios where all stochastic operators of interest, i.e.,~the set of operators $\{\hat{G}_k(\xi,\psi)\}_{k\in \mathcal{I}}$ (where $\mathcal{I}$ is some finite index set) are all Hermitian and mutually commuting. 
  That is, we consider the case where each stochastic operator of interest ($i\hat{G}_k(\xi,\psi)$) is anti-Hermitian, while $\hat{G}_k(\xi,\psi)$ itself is Hermitian and commutes with all other such operators, i.e.,~$\left[\hat{G}_k(\xi,\psi),\hat{G}_{k'}(\xi,\psi)\right]=0\, (\,\forall\,k,k'\in\mathcal{I}\,)$. Note that stochastic operators are not assumed to commute with the Hamiltonian itself (i.e., $\left[\hat{H},\hat{G}_k(\xi,\psi)\right]\neq0$ generically, for~$k\in\mathcal{I}$). 

Finally, we restrict our analysis to continuous stochastic operators and do not consider discrete jump processes in the quantum state.  That is, the Hilbert space dynamics under study is at least continuous for each stochastic trajectory of the quantum system under study; however, this does not imply any strict restriction on the noise variables, which may be both white (and discontinuous) or colored (and continuous). Concretely, the~quantum noise `$\xi$' denotes the various noise dependencies in the dynamics i.e.,~the entire set of all irreducible noise parameters, $\xi:=\,\{\xi_k\}_{k\in\mathcal{I}}$, which in the generic case are multiple noise variables, each dependent on time, which we suppress in our notation, $\xi_k:=\xi_k (t)$ for time $t>0$ and $k\in \mathcal{I}$.
Note that the index set of the noise and the index set of the stochastic operators are taken to be the same because we assume only linear coupling; that is, any stochastic operator with multiple noise contributions may be decomposed into operators $\hat{G}_k(\xi,\psi)$ that depend on $\xi$, but~only linearly. In~other words, terms of the form $\xi_k\,\xi_{k'}\,f(\psi)$ do not arise in the spectral decomposition of any $\hat{G}_k(\xi,\psi)$ (for $k,k'\in\mathcal{I}$ and any generic function $f$ in the Hilbert space).

Further, each $\xi_k$ is understood to satisfy its own equations of motions, which will be specified shortly. This has to be appropriately produced; otherwise, for ill-behaved noise processes, there may be no consistent notion of measurability or integration, while the driven quantum process may also evolve outside the Hilbert space. To~this end, we assume that each $\xi_k$ is a well-adapted Markovian stochastic process with an underlying natural filtered probability space $(\Omega,\mathcal{F},\mathcal{F}_t,\mathbb{Q})$~\cite{oksendal2003stochastic}. Here, $\Omega$ denotes the sample space encompassing all conceivable events of the stochastic process. $\mathcal{F}$, the~filtration, signifies the space encompassing all event collections (i.e., all measurable subsets of $\Omega$), and~$\mathbb{Q}$ represents a specified probability measure associating probabilities with abstract events in $\mathcal{F}$. The~overall probability complies with $\mathbb{Q}[\mathcal{F}]=1$. The~space $\mathcal{F}$ constitutes a $\sigma$-algebra. The~(natural) filtration, $\{\mathcal{F}_{0\leq t<\infty}\}\subset\mathcal{F}$, comprises a sequence of sub-$\sigma$-algebras with a causal ordering. It is dictated by all possible histories of a process leading up to time $t$. This sequence adheres to the causality condition $\mathcal{F}_{0}\subset\mathcal{F}_{t_1}\subset\mathcal{F}_{t_2}\,.\,.\,.\,\mathcal{F}_{s}\subset\mathcal{F}_{t}\subset\mathcal{F}$ for times, $0<t_1<t_2\,.\,.\,.\,s<t$, where each $\mathcal{F}_s$ represents the set of all event collections forming a history leading up to time $0<s<t$. 
 The~process $\xi_k$ is called an adapted process if $\xi_k$ is always $\mathcal{F}_t$ measurable, indicating that $\xi_k(t)$ possesses a well-defined probability for all possible histories leading up to time $t$, without~requiring knowledge of the future (non-anticipating and thus Markovian). All stochastic processes, $\xi$, considered below are thus assumed to be well adapted with an appropriate filtration. The~$\xi$ equations of motion will be specified shortly and are assumed to be Markovian, but each $\xi_k$ itself may possess temporal correlations and will control whether the driven quantum stochastic process itself is Markovian or~non-Markovian.

To avoid restricting our analysis to certain observables, such as in a Heisenberg picture approach~\cite{QFDR2,Araujo_2019_QFDR1,GardinerZoller2004}, we instead focus on the state ($\ket{\psi}\in \mathcal{H}$) dynamics, so that any information of interest, for any observable (not necessarily commuting with the dynamics) may be extractable at any time as per requirement. This is crucial, as~the behavior of different operators may be quite system-dependent and depend on their coupling to the noise. In~other words, in~a given quantum system, the time evolution of different operators may require different stochastic conventions (Ito or Stratonovich) to accurately describe effective Markovian dynamics, which, in principle, must be checked in each situation. 
Thus, it is a better strategy to consider instead the Schr\"odinger picture evolving the state, $\ket{\psi}$, and understand the nature of the corresponding stochastic convention to be applied to its effective Markovian dynamics. Then, all observable dynamics are also recovered in a consistent way.

\textbf{Quantum Stochastic Dynamics}: Following the above specifications, we now focus on analyzing stochastic equations of motion for the quantum state, $\ket{\psi}\in\mathcal{H}$, of~the form: 
 ($\hbar=1$):
\begin{align}i\frac{\partial}{\partial t}\ket{\psi}=\hat{H}\ket{\psi} + i\sum_{k\in\mathcal{I}}\hat{G}_k(\xi,\psi)\ket{\psi}.\label{Eq:1}\end{align}Note that $\ket{\psi}$ is a time-dependent state and if each $\hat{G}_k(\xi,\psi)$ is zero, the~above equation reduces to the Schr\"odinger equation, unitarily evolving closed quantum systems~\cite{Sakurai94}. Each $\hat{G}_k(\xi,\psi)$ modifies the unitary dynamics and injects stochasticity from source processes denoted by $\xi_k$. Since $\hat{G}_k(\xi,\psi)$ is non-linear, we note that although it is true that the equality holds, $\big[\hat{G}_k(\xi,\psi) + \hat{G}_j(\xi,\psi)\big]\ket{\psi}=\hat{G}_k(\xi,\psi)\ket{\psi} + \hat{G}_j(\xi,\psi)\ket{\psi}$ ($\forall j,k\in \mathcal{I};\,\forall\, \ket{\psi}\in\mathcal{H}$), in~general, for~$\ket{\psi}=\ket{\chi}+\ket{\phi}$, the expressions $\hat{G}_j(\xi,\psi)\ket{\psi}\neq\hat{G}_j(\xi,\chi)\ket{\chi}+\hat{G}_j(\xi,\phi)\ket{\phi}\,$ $ (\forall \ket{\psi},\ket{\phi},\ket{\chi} \in \mathcal{H}, \forall j\in \mathcal{I})$. We assume that all $\hat{G}_k(\xi,\psi)$ are (almost surely) bounded with respect to the Hilbert space operator norm (for all noise realizations); and as mentioned before, they are mutually commuting, and further, in~the non-finite case, their domains and target spaces are well specified and compatible with the Hamiltonian. In~general, the above equation, when driven by colored or temporally correlated noise, $\xi$, implies there is unspecified memory and is thus non-Markovian~\cite{Hanggi94,Luczka2005}. 

Said differently, without~further specification of the various $\xi$ processes, Equation~\eqref{Eq:1} is purely formal since the initial value problem is not defined; hence, knowledge of just the state $\ket{\psi}$ at some time is not sufficient to determine its time evolution, and thus Equation~\eqref{Eq:1} on~its own is generally non-Markovian~\cite{Hanggi94,Luczka2005}.
Given the above specifications in the non-Markovian regime, our aim in the Markovian limit is driven by physical viability, and hence, to realize CPTP dynamics, i.e.,~a quantum semi-group defined by its GKSL generator~\cite{GKS76,Lindblad1976}. In~the Markovian regime, the~form of $\hat{G}(\xi,\psi)$ is determined by constraints requiring a CPTP map in the context of open quantum systems or indeed fixing norm-preservation and forbidding superluminal signaling in the context of modified quantum theories~\cite{Gisin:1989sx,Bassi2015,aritro_FTL}. For~a single stochastic operator, these constraints ensure that under linear coupling with the noise, the~form of $\hat{G}(\xi,\psi)$ is given by the following~\cite{Gisin:1989sx,percival1998,Barchielli2009,GardinerZoller2004}:
\begin{align}
\hat{G}(\eta,\psi) := -\frac{\gamma^2}{2}\left(\hat{O}-\langle\hat{O}\rangle\right)^2 + \gamma\left(\hat{O}-\langle\hat{O}\rangle\right)\, \eta_t.\label{eq2:CSL_white_noise}
\end{align} Here, $\gamma$ is a constant, $\hat{O}$ is a Hermitian quantum operator, and the expression $\langle\hat{O}\rangle:=\bra{\psi_t}\hat{O}\ket{\psi_t}\hat{1}_\mathcal{H}$ denotes the usual quantum expectation at a particular time $t$, and $\hat{1}_\mathcal{H}$ is the identity operator on $\mathcal{H}$. We note that an additional locality constraint on $\hat{O}$ is required to forbid superluminal signaling in the case of tensor product Hilbert spaces modeling spatially separated parties~\cite{aritro_FTL}. The~principal feature of the Markovian equations is that the stochastic operator $\hat{G}(\xi,\psi)$ sources its randomness from Gaussian white noise~\cite{Hida1980}, $\xi_t\equiv\eta_t\in\mathbb{N}(0,1)$ (where $\mathbb{N}(0,1)$ is a standard normal distribution with unit variance), which is uncorrelated in time $\mathbb{E}_\eta[\eta_t\eta_s]=\delta(t-s)$, where the expectation value $\mathbb{E}_\eta[...]$ is taken over instantiations of $\eta_t$. As~a representative example, one may consider the case of decoherence in a qubit with $\hat{O}=\hat{\sigma}$, where $\hat{\sigma}$ denotes a Pauli spin~operator.
Note that the temporal integral of $\eta_t$ is the standard Wiener process or Brownian motion, $W_t=\int_0^t\eta_\tau \,d\tau$~\cite{oksendal2003stochastic,gardiner2004handbook} (this follows from Hida's classification of Brownian motion~\cite{Hida1980}). In~Ito's differential notation, this is denoted as $dW_t=\eta_t\,dt$. The~standard Wiener process~\cite{oksendal2003stochastic,gardiner2004handbook} is continuous with mean zero, $\mathbb{E}_\mathbb{Q}[W_t]=0$, and a~second moment $\mathbb{E}_\mathbb{Q}[W^2_t]=t$. Additionally, for~the differentials, we have $\mathbb{E}_\mathbb{Q}[dW_t]=0$ and $\mathbb{E}_\mathbb{Q}[dW_tdW_s]=0$ for $t\neq s$, alongside $\mathbb{E}_\mathbb{Q}[dW_t^2]=dt$, where $\mathbb{Q}$ denotes averaging with respect to the Wiener measure, which is also inherited by the well-adapted processes it drives. Note that these constitute the well-known Ito multiplication rules, given by $dt^2=dW_t dt=0$ and $dW_t^2=dt$.

Notice that, as before, $\hat{G}(\eta,\xi)$ is time-dependent due to its dependence on the state,= $\ket{\psi}$ at time point $t$ through the expectation values as well as due to its coupling to $\eta_t$. However, importantly, a~concrete stochastic convention has not yet been applied to Equation~\eqref{eq2:CSL_white_noise} yet, due to the presence of the non-linear multiplicative noise term ($\left[\hat{O}-\langle\hat{O}\rangle\right]\ket{\psi}\, \eta_t$). While naively using the Stratonovich convention on Equation~\eqref{eq2:CSL_white_noise} yields an unphysical, normalization-violating process with causality-violating features such as superluminal signaling (seen via non-linearities in the master equations~\cite{aritro_FTL}), the~same Equation~\eqref{eq2:CSL_white_noise} yields the well-known stochastic Schr\"odinger equation (SSE)~\cite{percival1998,Barchielli2009,GardinerZoller2004} only when written in Ito's convention using $dW_t=\eta_tdt$, thus yielding the following ($\hbar=1$):
\begin{align}
d\ket{\psi_t}&=-i\hat{H}\ket{\psi}\,dt\, + \hat{G}(\eta,\psi) \,\ket{\psi}\, dt\,; \nonumber\\
&=-i\hat{H}\ket{\psi}\,dt\, + \left[-\frac{\gamma^2}{2}\left(\hat{O}-\langle\hat{O}\rangle\right)^2 \,dt + \gamma\left(\hat{O}-\langle\hat{O}\rangle\right)\, dW_t\right] \,\ket{\psi}.\label{Eq3:white_ito_CSL}
\end{align}
The above equation is known to be an unraveling of a GKSL generator~\cite{Gisin84,Gisin:1989sx,percival1998}, showing that it yields a causal, Markovian, CPTP dynamical map. In~particular, the~(time-dependent) density operator, $\hat{\rho}_t$, defined as the statistical average of the Hilbert space trajectories obtained from Equation~\eqref{Eq3:white_ito_CSL}, i.e.,~$\hat{\rho_t}:=\mathbb{E}_{\mathbb{Q}}[\ket{\psi}\bra{\psi}]$, follows the GKSL master equations $\frac{\partial\hat{\rho}}{\partial t}=\mathcal{L}[ \hat{\rho}]$ with the standard generator, $\mathcal{L}[\cdot]:= -i[\hat{H},\cdot] + \gamma^2 \left(\hat{O}\,\cdot\,\hat{O}-\frac{1}{2}\{\hat{O}^2,\,\cdot\,\}\right)$~\cite{Gisin84,percival1998,GardinerZoller2004,aritro2,aritro3}. Such equations have been extensively studied in the theory of open quantum systems~\cite{GardinerZoller2004,Breur_Petr02} such as in continuous monitoring~\cite{Barchielli2009}, quantum Langevin models~\cite{QFDR2,Araujo_2019_QFDR1}, and the foundations of quantum theory~\cite{percival1998,Gisin:1989sx,Bassi_03_PhyRep,aritro_PhD}. In~a qubit (two level system) with $\hat{O}=\hat{\sigma}_z$, where $\hat{\sigma}_z$ denotes a Pauli spin matrix, the above equation takes the form of a standard decohering Lindblad master equation. Crucially, as~mentioned above, the~Stratonovich convention introduces non-linearities into the master equations, which generally imply that the resulting dynamical map is neither CPTP nor causal~\cite{aritro_FTL,Gisin:1989sx}. 
This stark distinction between the two conventions already indicates that a great deal of care must be taken to choose stochastic conventions. We will show that this choice of stochastic convention on the above white-noise-driven dynamics of Equation~\eqref{Eq3:white_ito_CSL} necessarily descends from a family of colored-noise-driven non-Markovian stochastic processes of the form of Equation~\eqref{Eq:1} on the Hilbert space for appropriately well-behaved noise processes to be explicitly classified shortly. This descent is shown constructively through a perturbative noise homogenization procedure~\cite{Pavliotis2008}, which functions as a temporal coarse-graining scheme while also ensuring consistent renormalization of both the noise and the state $\ket{\psi}$, such that all probabilistic interpretations remain well defined. Note that such a procedure constitutes a constructive generalization of the well-known Wong--Zakai theorems on random differential equations~\cite{WongZakaiReview,wongZakai1965convergence,WongZakai1969,WongZakai1965relation}, applied to quantum stochastic dynamics on Hilbert~spaces.

\section{Quantum Noise~Homogenization}
Having established our notation as well as the problem setting, we will now tackle this issue of stochastic conventions and show that it is not an ad hoc choice. In~this section, we coarse-grain the dynamics of Equation~\eqref{Eq:1} to yield generalizations of Equation~\eqref{Eq3:white_ito_CSL}. As~mentioned before, first, note that Equation~\eqref{Eq:1} in its given form is generically non-Markovian. This is simply because the~specifications of the noise are not provided. If~a memoryless specification is provided---for~example, from the previous section, $\xi\equiv\eta_t$, where $\xi$ has no temporal correlations (i.e., $\eta_t$ is a white noise process with no memory)---then Equation~\eqref{Eq:1} is rendered Markovian. In~all other non-trivial cases, Equation~\eqref{Eq:1} is generically non-Markovian. In~this situation, usually only the the statistical properties of $\xi$ are specified without its equations of motion or initial state, and thus, various approximations must be employed to close the dynamical equations of the system. For~instance, see the discussions in Refs.~\cite{Luczka2005, Hanggi94,Risken1996,Zwanzig2001_FDR} for an account of various approximation schemes within their regimes of~validity.

\textbf{Quantum~State-Noise~Augmentation:} In our case, we utilize the noise equations of motion, and thus, we begin with a convenient Markovian reformulation of Equation~\eqref{Eq:1}, which is obtained canonically as its maximal noise-augmented space. Simply put, instead of just considering the non-Markovian process of Equation~\eqref{Eq:1} evolving $\ket{\psi}$ with a dynamical law containing a partially specified $\xi$, we instead consider the larger space $\{\ket{\psi},\xi\}$ and specify the full dynamical system of equations therein. Thus, in addition to Equation~\eqref{Eq:1}, we specify the details of the noise equations of motion and consider the joint, noise-augmented dynamics of the quantum state:
\begin{align}
i\frac{\partial}{\partial t}\ket{\psi}&=\hat{H}\ket{\psi} + i\sum_{k\in\mathcal{I}}\hat{G}_k(\xi,\psi)\ket{\psi}.\notag \\
d\xi_k&=-f(\xi_k)\,\frac{dt}{\tau} + \, g(\xi_k)\frac{dW^k_t}{\sqrt{\tau}}.\label{Eq:4}\end{align}
Here, we specify that the noise $\xi$ is a continuous Markovian process, i.e.,~an Ito process where $f$ and $g$ are time-independent smooth real-valued functions and $\tau$ determines a time scale associated with the driving process, which is assumed to be much faster than time scales associated with the driven quantum state. We further assume that $\xi$ is centered, possesses a steady state, and~also has a corresponding white noise limit when $\tau\to0$~\cite{gardiner2004handbook,Pavliotis2008,Risken1996}. 
The above family of noise processes is deemed sufficiently general for our purposes. In~our context, since the noise $\xi$ is sourced from a physical environment, it is clearly important to consider Gaussian processes. 
The~only unique Gauss--Markov process upto rescaling is the Ornstein Uhlenbeck (OU) process with $\xi\in\mathbb{R}$ and $f(\xi_k)=\xi_k$, while $g(\xi_k):=\sqrt{2}$~\cite{OU_OG1930,gardiner2004handbook,Risken1996,Pavliotis2008}, which exhausts all Gaussian driving processes. We also further consider a non-Gaussian process termed spherical Brownian motion (SBM) with bounded $\xi\in[-1,1]$ with $f(\xi_k)=\xi_k$ and $g(\xi_k):=\sqrt{1-\xi_k^2}$~\cite{aritro2,aritro3,aritro_PhD,CNOSJacobiDEMNI2009518,CNOSPearsonDiffPaper2008,CNOStimelocalPearsontoJacobiAscione2021,CNOSWong1964}. Both the SBM and OU process possess steady states and exponentially decaying autocorrelations $\mathbb{E}[\xi_k(t)\xi_k(t')]\propto e^{-|t-t'|/\tau}\,\, (\,\forall\, k\in\mathcal{I}\,)$ where $\tau$ is the correlation time and both of these processes converge to white noise as $\tau\to0$~\cite{OU_OG1930,gardiner2004handbook,Risken1996,Pavliotis2008,CNOSWong1964,aritro_PhD}. 

\textbf{Temporal coarse-graining scheme:} Since we wish to finally converge on Equation~\eqref{Eq3:white_ito_CSL} and since the noise homogenization prescription allowing the temporal coarse-graining will be seen to not affect any explicitly non-stochastic term, we assume a decomposition into deterministic and stochastic parts; that is, each $\hat{G}_k(\xi,\psi)= A_k\,\hat{\mathcal{J}}_k(\psi) + B_k\,\hat{\mathcal{G}}_k(\xi,\psi)$, where $\mathcal{J}_k(\psi)$ is purely deterministic and $\hat{\mathcal{G}}_k(\xi,\psi):=\xi_k\hat{\Delta}_k$ is stochastic with quantum state-dependent operators $\hat{\Delta}_k:=\hat{\Delta}_k(\psi)$ (to be defined below) linearly coupled to the noise $\xi_k$. Further, $A_k$ and $B_k$ are real coupling constants. This does not lose generality since we have already assumed that the noise is linearly coupled to the stochastic operators, and further, it is well known that to avoid superluminal signaling, modifications of quantum dynamics must possess both deterministic and stochastic contributions, both non-linearly dependent on the state~\cite{Gisin:1989sx,Bassi2015,aritro_FTL}. 

We will now show that the precise family of non-Markovian quantum stochastic processes driven by colored noise (of the form of Equation~\eqref{Eq:4}), which converges to the form of the Markovian stochastic Schr\"odinger equation in Equation~\eqref{Eq3:white_ito_CSL} (with multiple stochastic operators), is given by: 
\begin{align}
d\ket{\psi} =  -i\hat{H}&\ket{\psi}dt+\sum_{k\in\mathcal{I}}\, \bigg[ -A_k \Big(\hat{\Delta}^2_k -\langle\hat{\Delta}^2_k\rangle\Big) +\,B_k\,\xi_k\hat{\Delta}_k \bigg]\ket{\psi}  dt, 
\label{Eq:App_Pair} \\
~~~\text{with},~~~ \hat{\Delta}_k &= \Big(\hat{O}_k -\langle\hat{O}_k\rangle\Big),\notag\\
\text{and},~~~ d\xi^k_t&=-f(\xi_k)\,\frac{dt}{\tau} + \, g(\xi_k)\frac{dW^k_t}{\sqrt{\tau}}.
\notag
\end{align}Here, note that all $\hat{O}_k$ ($k\in\mathcal{I}$) are mutually commuting and Hermitian, allowing a maximal commuting basis of the $\text{dim}\,\mathcal{H}=N$ Hilbert space, which we denote by $\{\ket{a}\}^{N}_{a=1}$. In~$N$-dimensions, the~corresponding time-dependent wave function in this basis is $\ket{\psi}:=\ket{\psi(t)}=\sum^N_{i=1}\psi_a\ket{a}$, where we have suppressed the time dependence in our notation. For~a qubit with $\text{dim}\,\mathcal{H}=2$, here, as~before, one may consider the case of $\hat{O}=\hat{\sigma}$ being that of a Pauli spin~operator. 

Since the operators $\hat{O}_k$ all have real eigenvalues, we will proceed with deriving the dynamics for the squared amplitudes $z_a := |\bra{a}\psi\rangle|^2$.  From~Equation~\eqref{Eq:App_Pair}, the~dynamical equations for the components $\psi_a$ are given by:
\begin{align}
d\psi_a = -i \langle a|\hat{H}\ket{\psi}dt+\psi_a \sum_{k\in\mathcal{I}}\Biggl[-A_k  \Bigl( \langle a|\hat{\Delta}^2_k|a\rangle-\langle\hat{\Delta}^2_k\rangle\Bigr)  + B_k\,\xi_k\,\Bigl(\langle a|\hat{O}_k|a\rangle-\langle\hat{O}_k\rangle\Bigr)\Biggr]dt.
\end{align}Note that for~an arbitrary smooth function $f(\psi)$, the~temporal integral of the form $\int f(\psi)\xi_k\,dt$ is a regular Riemann integral, since the $d\psi_a$ process has a vanishing quadratic variation ($d\psi_a^2=0$). This is because the noise $\xi_k$ is continuous while its integral is continuous and once-differentiable. Thus, there is no difference between the Ito and Stratonovich descriptions in this case of colored noise driving~\cite{Hanggi94,Ito_strat1,aritro2}, and~using the regular rules of calculus, we obtain the evolution equations for $dz_a=d(\psi^*_a\psi_a)$ given by:
\begin{align}
    dz_a&=(H_a+J_a)\,dt + \,\sum_{k\in\mathcal{I}}\, G_{ak}\xi_k\,dt.
\end{align}
Here, we define $H_a=i(\psi_a\bra{a}\hat{H}\ket{\psi}^*-\psi^*_a\bra{a}\hat{H}\ket{\psi})$ as the contribution from the Hamiltonian, which is real-valued, and $J_a = -2 z_a \sum_{k\in \mathcal{I}}\, A_k \left( \langle a|\hat{\Delta}^2_k|a\rangle-\langle\hat{\Delta}^2_k\rangle\right)$, and the coefficient of the stochastic part, $G_{ak} = 2 B_k\,z_a\,\left(\langle a|\hat{O}_k|a\rangle-\langle\hat{O}_k\rangle\right)$. 

Now, we shall coarse-grain the colored-noise-driven dynamics to obtain effective Markovian dynamics through a multi-scale noise homogenization procedure. To~perform the noise homogenization, the~first step is to isolate a factor of the fast correlation time of the noise $\sqrt{\tau}$ from  $G_{ak}$ by rescaling $G_{ak}\to \sqrt{\tau}G_{ak}$ and $dt\to dt/\sqrt{\tau}$. This effectively initializes the coarse graining of the dynamics, since the stochastic differential equations are formally understood as discrete sums over temporal partitions, which increase as $\frac{dt}{\sqrt{\tau}}>dt$ as $\tau\to0$ for a fixed partition size. Hence, we consider the limit $\tau\rightarrow0$ in such a way that $B_k^2\tau$ remains finite, thus executing a temporal coarse graining with noise homogenization~\cite{Pavliotis2008}. The~complete rescaled system of equations is now given by:
\begin{align}
    dz_a&=(H_a+J_a)\,dt + \,\sum_{k\in\mathcal{I}}\, G_{ak}\xi_k\,\frac{dt}{\sqrt{\tau}};
   \notag \\
   &\text{with,}~~~\notag\\
    G_{ak}&=2\,z_a\,\sqrt{\tau B_k^2}\,\left(\langle a|\hat{O}_k|a\rangle-\langle\hat{O}_k\rangle\right), \notag \\
   d\xi_k&=-f(\xi_k)\,\frac{dt}{\tau} + g(\xi_k)\frac{dW^k_t}{\sqrt{\tau}}.
   \label{Eq:ZPair}
\end{align}Here, $a\in [1,N]$ denotes the Hilbert space basis, while $k\in\mathcal{I}$ labels the various stochastic operators. To~extract the limit for $\tau\rightarrow0$, we employ the Kolmogorov backward equation (KB)~\cite{oksendal2003stochastic,gardiner2004handbook,Risken1996}, which is obtained from the generator, $\Lambda(\mathbf{z},\bm{\xi})$, of~the joint stochastic process in the augmented quantum state-noise space, relabeled as $\{\mathbf{z},\bm{\xi}\}:=\{\ket{\psi},\xi\}$ in Equation~\eqref{Eq:ZPair}. 

It is now evident that considering the joint statistics in the augmented space $\{\mathbf{z},\bm{\xi}\}$ is a necessary step to obtain Markovian dynamics for the colored-noise-driven dynamics, even in a simple model where the noise is independent of the quantum state dynamics but the state dynamics itself, depends on the noise. Thus, the~expressions of the quantum state $\mathbf{z}(t)$ depend on \textcolor{black}{$\bm{\xi}(t)=\bm{\xi}(0)+\int_0^t\,d\bm{\xi}(t)$}, which requires information of $\bm{\xi}(t)$ up to time point $t$, rendering $\mathbf{z}(t)$ on its own non-Markovian without this specification. 
Augmenting the $\mathbf{z}(t)$ with the $\bm{\xi}(t)$ process, however, renders the combination Markovian~\cite{Hanggi94,Luczka2005,aritro2}, and by exploiting this, we may proceed directly to the corresponding Kolmogorov backward equations~\cite{oksendal2003stochastic,Pavliotis2008,Risken1996,gardiner2004handbook} for the system in Equation~\eqref{Eq:ZPair}, which yields the evolution of likelihood densities $\rho(\mathbf{z},\bm{\xi},t)$ in the higher-dimensional augmented space and has the form:
\begin{align}
 -\partial_t\rho(\mathbf{z},\bm{\xi},t) =& \Lambda(\mathbf{z},\bm{\xi})\,\rho(\mathbf{z},\bm{\xi},t); \\
    \text{where,} ~~~~
    \Lambda(\mathbf{z},\bm{\xi}) :=& \Lambda_H^J(\mathbf{z}) + \frac{1}{\sqrt{\tau}}\Lambda_{G}(\mathbf{z},\bm{\xi}) + \frac{1}{\tau}\Lambda_\xi(\bm{\xi});\nonumber \\
    \text{and,} ~~~~~~
    \Lambda_H^J(\mathbf{z}) ~=&~\sum^N_{a=1}\,(H_a+J_a)\,\frac{\partial}{\partial\,z_a}, \nonumber\\
    \Lambda_G(\mathbf{z},\bm{\xi})  ~=&~ \sum_{a=1}^{N}\sum_{k\in\mathcal{I}}\,G_{ak}\,\xi_k\,\frac{\partial}{\partial\,z_a},\nonumber\\
    \Lambda_\xi(\bm{\xi}) ~=&~ \sum_{k\in\mathcal{I}}\,\left(-\xi_k \frac{\partial}{\partial{\xi_k}}+\frac{g^2(\xi_k)}{2}\frac{\partial^2}{\partial\xi^2_k}\right).\label{Eq:Op defs BK}
\end{align}
Note that the characteristic operator $\Lambda(\mathbf{z},\bm{\xi},t)$ is a function of variables on the domains of the stochastic processes $z_a(t)$ and $\xi_k(t)$, and~in Equation~\eqref{Eq:Op defs BK}, they are interpreted directly as functions of $z_a$ and $\xi_k$ themselves (rather than of the stochastic processes $z_a(t)$ and $\xi_k(t)$; thus each stochastic variable constitutes a dimension in the Kolmogorov system). The~differential operator $\Lambda_H^J$ depends on $\mathbf{z}$ alone, while $\Lambda_G$ depends on $\mathbf{z}$ and $\bm{\xi}$, and~$\Lambda_\xi$ depends on $\bm{\xi}$ alone, and~these dependencies are dropped for ease of notation from here on. Notice that the steady-state distributions $\rho^{\infty}(\bm{\xi}):=\prod_k\rho^\infty(\xi_k)$ (also termed the invariant or null subspace) are given by the adjoint generator action $\Lambda^{*}_\xi\rho^{\infty}(\bm{\xi})=0$, which is equivalent to our previous assumption that the (forward) Fokker--Planck--Kolmogorov system of the noise, $\bm{\xi}$, has a long time steady-state distribution $\rho^{\infty}(\bm{\xi})$.
In the limit of small $\tau$, the~coefficient of $\Lambda_\xi$ guarantees that the noise dynamics is always faster than that of the state. Notice, however, that we will not assume the noise to always remain in its steady or late-time distribution $\rho^{\infty}(\bm{\xi})$, since assuming so would result in a straightforward, noise-free scenario that fails to account for the time-dependent nature of the dynamics induced by noise in the probability density~\cite{Luczka2005,HorsthemkeBook2006,Pavliotis2008}. Note that the steady-state distribution exists for both the OU ($\rho^{\infty}(\xi)=\frac{1}{\sqrt{2\pi}}e^{\frac{-\xi^2}{2}}$) and SBM ($\rho^{\infty}(\xi)=\frac{1}{2}$) processes. 

The Kolmogorov backward system in Equation~\eqref{Eq:Op defs BK} offer a methodical way to treat the density $\rho(\mathbf{z},\bm{\xi},t)$ by representing it as a perturbative expansion in $\tau^{1/2}$, treating the characteristic time of the noise as a small parameter:
\begin{align}
\rho(\mathbf{z},\bm{\xi},t)= \sum^{\infty}_{k=0} \tau^{k/2} \rho_k = \rho_0 + \sqrt{\tau}\rho_1 + \mathcal{O}[\tau].
\label{Eq:rho_tau_exp}
\end{align}
Utilizing Equation~\eqref{Eq:Op defs BK}, we derive distinct equations for each power of $\tau$:
\begin{align}
\tau^{-1}      &:\quad -\Lambda_{\xi}\rho_0 ~~~=~~ 0\,; \notag\\[4pt]
\tau^{-1/2}    &:\quad -\Lambda_{\xi}\rho_1 ~~~=~~ \Lambda_G\rho_0\,; \notag\\[4pt]
\tau^{0}       &:\quad -\Lambda_{\xi}\rho_2~~~ = ~(\Lambda_H^J + \partial_t)\rho_0 + \Lambda_G\rho_1\,; \notag\\[4pt]
\tau^{p/2}     &:\quad -\Lambda_{\xi}\rho_{p+2} = (\Lambda_H^J + \partial_t)\rho_p + \Lambda_G\rho_{p+1}\,.
\label{Eq:Pert4}
\end{align}
The expression on the final line holds for $p>-2$, with~the condition that $\rho_{p}=0$ for all ${p<0}$. Notice that because $\Lambda_{\xi}$ is a differential operator in $\bm{\xi}$ only, the~first line implies that $\rho_0$ depends only on the quantum state and time, denoted as $\rho_0 := \rho_0(\mathbf{z},t)$. The~subsequent set of equations comprises expressions of the form $\Lambda_{\xi}\rho_k=f_k(\mathbf{z},\bm{\xi})$. Solutions to these can be constructed using Fredholm's alternative theorem~\cite{Pavliotis2008,HorsthemkeBook2006,BoninTraversaSmallTau}. In~a finite-dimensional Hilbert space, this theorem states that an operator equation $A {x_i}= {y_i}$ with operator $A$ and vectors ${x_i}$ and ${y_i}$ has a solution if $n\cdot y_i=0$ for vectors $n$ in the null subspace of the adjoint operator $A^*$ (i.e., $A^{*}{n}=0$). The~solvability condition ${n}\cdot {y_i}=0~\forall i$ can be utilized to construct the solution for the system of equations. This is accomplished by repeated operations of $x_i=A^{-1}\,y_i$, which guarantees a solution if $y_i$ is orthogonal to the null~subspace.

This method applies to expressions like $\Lambda_{\xi}\rho_k=f_k(\mathbf{z},\bm{\xi})$ in Eq.~\eqref{Eq:Pert4} because the adjoint operator $\Lambda_{\xi}^*$ describes the forward evolution in Fokker--Planck--Kolmogorov equations of the noise alone~(see the previous sections and Refs.~\cite{Risken1996,gardiner2004handbook,HorsthemkeBook2006,Pavliotis2008}). As~both SBM and OU processes allow long time steady-state probability distributions, we have $\Lambda_{\xi}^{*}\rho^{\infty}(\bm{\xi})=0$, defining the null subspace of the adjoint operator. Further, the~orthogonality of a generic function $y(\mathbf{z},\bm{\xi})$ with the null subspace is expressed through the (function space) inner product: $\mathbb{E}^{\infty}_{\xi}[y(\mathbf{z},\bm{\xi})]:=\int d\bm{\xi}\,\rho^\infty(\bm{\xi}) y(\bm{z},\bm{\xi}) =0$. With~this, the~solvability condition for the $O(\tau^{-1/2})$ equation becomes $\mathbb{E}^{\infty}_{\xi}[\Lambda_G\rho_0]=0$. This condition, termed the centering condition, characterizes the coupling between the state dynamics and the noise~\cite{Pavliotis2008,HorsthemkeBook2006,BoninTraversaSmallTau}. For~the OU and SBM processes, their ergodicity properties and the linear coupling (in this case, a separate $\xi_k$ for each $G_{ak}$) guarantee that the centering condition is satisfied. To~see this, note that the form of $\Lambda_G$ has a differential operator only in $\mathbf{z}$, and since  $\mathbb{E}^{\infty}_{\xi}[\xi]=0$, we trivially have $\mathbb{E}^{\infty}_{\xi}[\Lambda_G\rho_0]=0$.  Thus, a~solution for the first-order component $\rho_1$ exists, given $\rho_0$, as~$\rho_1 = - \Lambda^{-1}_{\xi}\Lambda_G\rho_0$. 

Substituting this $\rho_1$ expression into the equation for $\Lambda_{\xi}\rho_2$ yields the subsequent solvability condition:
\begin{align}
    \mathbb{E}^{\infty}_{\xi}\left[ (\Lambda_H^J+\partial_t)\rho_0 -\Lambda_G\Lambda^{-1}_{\xi}\Lambda_G\rho_0 \right]=0.
    \label{Eq:Final_Solvbl0}
\end{align}
Only the final term on the right-hand side relies on $\xi$, which simplifies this expression to:
\begin{align}
    (\Lambda_H^J+\partial_t)\rho_0 -\mathbb{E}^{\infty}_{\xi}\left[\Lambda_G\Lambda^{-1}_{\xi}\Lambda_G\rho_0 \right]=0.
    \label{Eq:Final_Solvbl}
\end{align}
The remaining expectation value is computed using the so-called cell problem ansatz~\cite{Pavliotis2008,HorsthemkeBook2006,BoninTraversaSmallTau}. This begins with noticing that $\Lambda_G = \sum_{a,k} \xi_k \,G_{ak}\frac{\partial}{\partial\,z_a}$, which makes $\Lambda^{-1}_{\xi}\Lambda_G\rho_0$ expressible as $\Lambda^{-1}_{\xi} (\sum_k\xi_k F_k(\mathbf{z}))$ for an arbitrary function $F_k(\mathbf{z})$. The~cell problem ansatz then posits the existence of a function $\Phi(\mathbf{z},\bm{\xi})$ such that $\Lambda^{-1}_\xi\,\Lambda_G\rho_0 = \Phi(\mathbf{z},\bm{\xi}) $. If~this function exists, it implies $\Lambda_{\xi} \Phi(\mathbf{z},\bm{\xi}) = \Lambda_G\rho_0$. Finding such a function $\Phi(\mathbf{z},\bm{\xi})$ is equivalent to determining an expression for $\Lambda^{-1}_{\xi}\Lambda_G\rho_0$, which is required for evaluating the solvability condition. For~our case, using $\Lambda_{\xi}$, the~function, $$\Phi(\mathbf{z},\bm{\xi}) = -\sum^N_{a=1}\sum_{k\in\mathcal{I}}\xi_k\,G_{ak}\frac{\partial\,\rho_0}{\partial\,z_a},$$ solves the cell problem for both the OU and SBM noise processes. Now, since, $\Lambda_G\Lambda^{-1}_\xi\,\Lambda_G\rho_0 = \Lambda_G\Phi(\mathbf{z},\bm{\xi})$, we may evaluate the solvability condition in Equation~\eqref{Eq:Final_Solvbl} by substituting this into the expectation value, which yields:
\begin{align}
\mathbb{E}^{\infty}_{\xi}\left[\Lambda_G\Lambda^{-1}_{\xi}\Lambda_G\rho_0 \right]=-\sum_{a,b=1}^N\sum_{k\in \mathcal{I}} \mathbb{E}^{\infty}_{\xi}\left[ \xi^2_k\right]\,\Bigg( G_{ak}\frac{\partial G_{bk}}{\partial\,z_a}\,\frac{\partial\rho_0}{\partial\,z_b} +  G_{ak}\,G_{bk}\frac{\partial^2\rho_0}{\partial\,z_a\,\partial\,z_b}\,\Bigg).
\label{Eq:Final_Solvbl2}
\end{align}
Here, we use our assumption that all $\xi_k$ are stochastic processes that are uncorrelated with each other, $\mathbb{E}^{\infty}_{\xi}\left[ \xi_k\xi_{k'}\right]=0$. Together with Equation~\eqref{Eq:Final_Solvbl}, Equation~\eqref{Eq:Final_Solvbl2} establishes the solvability condition for dynamics up to order $\tau^0$. This homogenizes over the stochastic variables to order $\tau^0$, providing an expression for the time evolution of probabilities for the quantum state $\mathbf{z}(t)$ alone, which is given by:
\begin{align}
&\left[\partial_t + \Lambda_H^J +\tilde{\Lambda}_G \right] \rho_0(\mathbf{z},t) =0;\label{eq:stratBK}\\
\text{with,}~~~\tilde{\Lambda}_G&:=\frac{1}{2}\sum_{a,b=1}^N\sum_{k\in\mathcal{I}} \Bigg( \tilde{G}_{ak}\frac{\partial \tilde{G}_{bk}}{\partial\,z_a}\,\frac{\partial\rho_0}{\partial\,z_b} +  \tilde{G}_{ak}\,\tilde{G}_{bk}\frac{\partial^2\rho_0}{\partial\,z_a\partial\,z_b}\,\Bigg).
\notag
\end{align}
Here, we rescaled $G_{ak}$ by introducing $\tilde{G}_{ak}:=2z_a\sqrt{\mathcal{D}_k}\,\left(\langle a|\hat{O}_k|a\rangle-\langle\hat{O}_k\rangle\right)
$, with~the effective coupling defined as $\mathcal{D}_k=2\mathbb{E}^{\infty}_{\xi}\left[ \xi^2_k\right]B_k^2\tau $. This equation signifies the solvability condition for the system of Equation~\eqref{Eq:ZPair}, and~the solutions of Equation~\eqref{eq:stratBK} weakly correspond (agreeing at the level of ensemble averages) to those of Equation~\eqref{Eq:ZPair} in the $\tau\to 0$ limit, when the distribution $\rho(\mathbf{z},\bm{\xi},t)$ equals $\rho_0(\mathbf{z},t)$. Furthermore, Equation~\eqref{eq:stratBK} represents a Kolmogorov backward equation (in the Stratonovich representation) for the time evolution of the likelihoods of $\mathbf{z}(t)$ alone (after homogenizing over the noise)~\cite{gardiner2004handbook,Hanggi94,oksendal2003stochastic}. Notably, this equation straightforwardly coincides with the Stratonovich white-noise-driven process~\cite{Pavliotis2008,HorsthemkeBook2006,BoninTraversaSmallTau, gardiner2004handbook,oksendal2003stochastic}:
\begin{align}
    d\,z_a = (H_a+J_a)\,dt + \sum_{k\in\mathcal{I}}\tilde{G}_{ak}\circ\,dW^k_t\,+\mathcal{O}[\tau].
\end{align}
Because the usual rules of calculus apply in the Stratonovich representation (where $\circ$ denotes the Stratonovich product), the~quantum stochastic dynamics on the Hilbert space for the original pair of processes involving both $\ket{\psi}$ and $\{\xi_k\}_{k\in \mathcal{I}}$ can now be represented in the $\tau\rightarrow 0$ limit by a single effective quantum stochastic process on the Hilbert space:
\begin{align}
    d\ket{\psi} = -i\hat{H}\ket{\psi}\,dt+\sum_k\Biggl[  -A_k\bigg(\hat{\Delta}^2_k -\langle\hat{\Delta}_k ^2\rangle\bigg)\ket{\psi}\,dt+\,\sqrt{\mathcal{D}_k}\,\bigg(\hat{O}_k -\langle\hat{O}_k\rangle\bigg)\ket{\psi}\circ dW_t^k\Biggr].
    \label{Eq:WhiteNoiseFinal}
\end{align}
This establishes the existence of an analytically tractable Markovian limit in the joint state-noise dynamics of Equation~\eqref{Eq:App_Pair}, such that in the limit of $\tau\rightarrow0$, the~dynamics is weakly equivalent to the white-noise-driven process of Equation~\eqref{Eq:WhiteNoiseFinal}. This is an expected result, in~light of the Wong--Zakai theorems~\cite{WongZakaiReview,wongZakai1965convergence,WongZakai1969,WongZakai1965relation}, and~can be generalized~further. 

In our situation of interest, the~Markovian limit (for $\tau\rightarrow0$) is thus found to follow from the straightforward prescription:
\begin{align}
    \lim_{\tau\to0} \, \int^t_0 \,B_k\,\Bigl(\hat{O}_k-\langle\hat{O}_k\rangle\Bigr) \,|\psi\rangle \,\xi^k_t \,dt ~~ \to ~~  \,\int_0^t \,\tilde{B}_k\,\Bigl(\hat{O}_k-\langle\hat{O}_k\rangle\Bigr) \,|\psi\rangle \circ dW^k_t.\label{Eq:Col_to_Strat}
\end{align}
The above prescription clearly shows that colored-noise-driven non-Markovian quantum stochastic processes, after~coarse-graining, converge upon Markovian quantum stochastic processes driven by white noise but in the Stratonovich convention and with a renormalized diffusion coefficient, $\tilde{B}_k=\sqrt{\mathcal{D}_k}=B_k\sqrt{2\mathbb{E}^{\infty}_{\xi}\left[ \xi^2_k\right]\tau}$. The~conversion of the above Stratonovich processes to their corresponding Ito representation hence leads to a further correction term of order $dt$ as discussed~below.

We may convert Equation~\eqref{Eq:WhiteNoiseFinal} from the Stratonovich representation to the Ito representation using $\,X\circ dY= \,X\,dY\,+\frac{1}{2}dX \,dY$~\cite{oksendal2003stochastic} with $X=\,\sqrt{\mathcal{D}_k}\,\Big(\hat{O}_k -\langle\hat{O}_k\rangle\Big)\ket{\psi}$ and $dY=dW^k_t$ for each $k\in\mathcal{I}$, resulting in a correction term of order $dt$ being added to Equation~\eqref{Eq:WhiteNoiseFinal} in the Ito convention, defined as $\hat{C}\ket{\psi}dt:=\frac{1}{2}dX \,dY$. The~Stratonovich correction then simplifies to:
\begin{align}
\hat{C}\ket{\psi}dt&=\sum_{k\in\mathcal{I}}\mathcal{D}_k\bigg(\frac{1}{2} 
    \left[\hat{{O}}_k - \langle\hat{O}_k\rangle\right]^2 
    -  \left[\langle\hat{O}_k^2\rangle - \langle\hat{O}_k\rangle^2\right]\bigg )\ket{\psi}dt\notag\\&=\sum_{k\in\mathcal{I}}\mathcal{D}_k\bigg(\frac{1}{2} 
    \hat{\Delta}_k ^2 
    -  \langle\hat{\Delta}_k ^2\rangle \bigg )\ket{\psi}dt.
\end{align}
Further, to~achieve norm preservation, one must crucially impose a manner of fluctuation dissipation relation~\cite{aritro_PhD,aritro2,aritro3} between the deterministic coefficient $A_k$ and the diffusion coefficient $\sqrt{\mathcal{D}_k}$ of the form $A_k=\mathcal{D}_k=\gamma_k^2$, which then results in Equation~\eqref{Eq:WhiteNoiseFinal} taking the form of the well-known stochastic Schr\"odinger equation (Equation~\eqref{Eq3:white_ito_CSL}), which is a Markovian, norm preserving quantum stochastic process driven by white noise and \textcolor{black}{possesses the necessary structure to unravel linear CPTP dynamics without causality violations}~\cite{aritro_FTL,Gisin:1989sx,Bassi2015}, given by:
\begin{align}
d\ket{\psi_t}&=-i\hat{H}\ket{\psi}\,dt\, +\sum_{k\in\mathcal{I}} \Biggl[-\frac{\gamma_k^2}{2}\Bigl(\hat{O}_k-\langle\hat{O}_k\rangle\Bigr)^2 \,dt + \gamma_k\Bigl(\hat{O}_k-\langle\hat{O}_k\rangle\Bigr)\, dW^k_t\Biggr] \,\ket{\psi}.\label{Eq3:white_ito_final}
\end{align}
Thus, we have shown that the SSE of the form of Equation~\eqref{Eq3:white_ito_CSL} and Equation~\eqref{Eq3:white_ito_final} may be recovered as the Markovian limit of the generally non-Markovian quantum stochastic process described by Equation~\eqref{Eq:1} driven by colored noise. The quantum noise homogenization procedure links these two limits, and further requiring causal CPTP Markovian dynamics enforces a choice based on physical admissibility and removes the ambiguity between the Ito and Stratonovich conventions.
\section{Conclusions}
In this section, we summarize the main analytic results obtained by applying the quantum noise homogenization scheme to the augmented state-noise dynamics of Equation~\eqref{Eq:App_Pair} and conclude by highlighting its consequences for physically relevant scenarios. Firstly, under~the assumptions stated in Section~\ref{sec:2}, we show that the expansion of the joint Kolmogorov (backward) system of equations in Equation~\eqref{Eq:rho_tau_exp} admits a regular perturbative solution in powers of relevant noise time scales that allow for a quantum noise homogenization procedure, connecting colored-noise-driven, non-Markovian quantum stochastic processes to their coarse Markovian counterparts. The~order by order solvability conditions then yield the homogenized Markovian backward operator (Equation~\eqref{eq:stratBK}) for the quantum state alone. 
 Indeed, for~driving noise processes of the form of Equation~\eqref{Eq:4} (including OU and SBM) possessing a white noise limit, the~non-Markovian colored-noise driven quantum dynamics are seen to be weakly equivalent (i.e., equivalent at the level of ensemble probabilities) to a white-noise-driven, Markovian, Stratonovich quantum stochastic process (Equation~\eqref{Eq:WhiteNoiseFinal}), with renormalized noise couplings as well as deterministic~corrections in Ito’s convention.

Equation~\eqref{Eq:Col_to_Strat} is the central constructive realization of our work and shows a Wong--Zakai-type limit for non-Markovian quantum stochastic processes driven by colored noise, wherein the colored multiplicative noise coupling maps onto a Stratonovich white noise coupling in the limit of small characteristic time scales of the noise. Further, the conversion of the Stratonovich process to the Ito convention produces a deterministic correction term. In~our setting, this precise correction \textcolor{black}{and the restriction to CPTP dynamics} allows a fluctuation--dissipation relation to be imposed, which guarantees norm preservation and yields the well-known Ito form of the stochastic Schr\"odinger equation that unravels linear GKSL generators, which precludes causality violations. Thus, the precise Ito form of the stochastic Schr\"odinger equation (Equation~\eqref{Eq3:white_ito_final}), well known in the literature, appears only after adding the deterministic drift to the Stratonovich limit, which itself descends from the colored-noise-driven dynamics of Equation~\eqref{Eq:App_Pair}.

\textcolor{black}{Thus, we have demonstrated that the standard Markovian stochastic Schr\"odinger equations, which unravel CPTP dynamics, employed in many open quantum system models as well as in foundations of quantum theory, emerge naturally from the quantum noise homogenization scheme when applied to a broad class of non-Markovian models driven by colored noise,~particularly those with correlated Gaussian noise, a~scenario of central physical relevance. This result resolves the Ito--Stratonovich ambiguity for quantum stochastic processes whose Markovian limit processes unravel causal CPTP dynamics. }

\textcolor{black}{Specifically, we showed that the Markovian (white noise) limit of potentially non-linear quantum stochastic processes driven by temporally correlated multiplicative noise should be formulated using the Stratonovich convention with appropriately renormalized stochastic coefficients. These renormalized stochastic coefficients function as effective diffusion constants and also contribute to additional drift corrections in Ito's convention. These coefficients must be computed in each case using the quantum noise homogenization procedure presented in this article. Finally, our results show that the relevant class of Markovian quantum stochastic processes driven by white noise, which are unravellings of linear CPTP dynamics or Lindblad master equations, may be obtained as appropriate Markovian limits of a broad class of non-Markovian quantum stochastic processes driven by colored noise  (adhering to the assumptions in Section~\ref{sec:2}), which are described in Equations~\eqref{Eq:4} and~\eqref{Eq:App_Pair}.}

Several research directions towards generalizing this scheme follow naturally, which are left for future investigations: considering the (separable) infinite-dimensional case; extending our method to non-commuting and generic non-Hermitian stochastic operators, as~well as to the assessment of operator-ordering subtleties in the homogenized limit; and extending our method to a more general class of driving noise processes as well as limits wherein the relevant time scales of the noise vary significantly. \textcolor{black}{Further, our scheme considers quantum stochastic processes whose Markovian limits unravel CPTP dynamics; however, it is known that it may not exhaust all physically relevant scenarios (see Refs.~\cite{NoCP_ShajiSudarshan2005,NoCP_Pechukas1994,NoCP_Alicki1995,NoCP_Pechukas1995Reply}). Indeed, our restriction to CPTP dynamics enables the identification of the fluctuation dissipation relation and enforcing causality, which in turn allows a choice of physically admissible processes and resolves the Ito--Stratonovich ambiguity for a large class of models (subject to the assumptions stated in Section~\ref{sec:2}) self-consistently. Although~the quantum noise homogenization can be applied to any stochastic operator (adhering to the aforementioned assumptions), how the entire scheme can be consistently extended to the non-CP case remains an open~problem.}

In conclusion, we have formulated an analytically closable and controlled quantum noise homogenization procedure that connects a wide class of non-Markovian, colored-noise-driven quantum stochastic processes to their effective Markovian limit process, unraveling causal, CPTP dynamics, and~in doing so, we resolved the operational ambiguity between the Ito and Stratonovich conventions for a large class of physically relevant quantum stochastic processes. We expect the present framework to be broadly useful wherever colored multiplicative noise appears in quantum trajectories of open quantum systems, collapse models, and~noise-driven many-body systems, by~providing a concrete algorithm for choosing the physically relevant stochastic~calculus.
\\ \\
\emph{\textbf{Acknowledgments}} ---
I acknowledge J\~nanananda Seva Sangha and S.M.P Devi for their gracious support and useful discussions.

\end{document}